\documentstyle[pre,aps,epsf]{revtex}
\begin{document} 
\draft 
\title{ 
Fluctuations in confined nematic liquid crystals\\
 in a regime of critical wetting }
 
\author{N. B. Ivanov }
\address{Institute  for Solid  State 
Physics, Bulgarian Academy of Sciences,\\ 
Tzarigradsko 
chaussee-72, 1784 Sofia, Bulgaria} 
 
\date{6 December, 1998}
\maketitle 
\begin{abstract} 
Within the  macroscopic Landau-de Gennes approach, we examine
the Gaussian normal mode fluctuations  of  semi-infinite
nematic liquid crystals in a regime of critical wetting.
It is argued that surface free-energy
potentials which strongly suppress
the long-range nematic order favor the appearance
of bound biaxial nematic-director fluctuation modes,
located in the domain occupied by the thermodynamic phase
wetting the wall. Instead, substrates
enhancing the orientational order promote
the existence of
uniaxial nematic-director local excitations. Close to the phase
coexistence temperature both types of local exciations are strongly
softened as compared to their
bulk counterparts and acquire characteristic
cusplike low-energy spectra. These spectrum peculiarities
are directly connected to the critical behavior
of the mean-field interface position
and can  provide a valuable insight
on the nature of surface interactions and
critical wetting phenomena
in nematic liquid crystals.
Possible changes in the local director mode properties
resulting from the critical interface position
fluctuations and  order electricity effects are also discussed.
\end{abstract} 
\section{Introduction}
Liquid crystals in confined geometries (such as plates, droplets, porous
glasses, and polymer networks)
constitute an interesting
class of thermodynamic systems both  from experimental and
theoretical viewpoints \cite{crawford,chakrabarti,ramdane,vandenbrouck}.
In particular, the generic critical
wetting phenomena \cite{pandit,dietrich},
studied in a large number of scalar order-parameter systems,
are expected to acquire some
new interesting feartures near the bulk nematic-isotropic coexistence
temperature $T_{NI}$ arising from  the complicated nematic
order parameter (a symmetric traceless second-rank tensor).
In confined nematic liquid crystals the critical wetting  transition
usually appears as a {\em continuous interface
delocalization
transition} at bulk coexistence, when the nematic-isotropic interface
critically intervenes into the bulk as $T_{NI}$ is approached.
Typical features of a critical interface
delocalization transition are the
diverging interface position and the existence
of a {\em soft fluctuation mode} directly connected to the interface
position fluctuations (for a review see, e.g.,
Ref. \onlinecite{lipowsky}).
At a mean-field level and in a uniform-director approximation,
the macroscopic Landau-de Gennes approach\cite{degennes}
reproduces the well-known  critical wetting results originally
established for scalar order-parameter systems, so that
many known results can be directly applied to nematic liquid
crystals (for a review on wetting transitions in liquid
crystals see Ref. \onlinecite{crawford1} and references therein).

Concerning the fluctuation dynamics, the situation in
systems with an orientational rong-range order
might be substantionally changed. Generally,
the  nematic order-parameter fluctuations
are characterized by five independent fluctuation fields, one of them
being related to the scalar order-parameter fluctuations, whereas
the other four fields describe the transverse
fluctuations of the nematic director.
Therefore, an interesting issue is to examine
nematic director fluctuations
in a regime of critical wetting, as one can expect
that some peculiarities of normal-mode director fluctuations
will reflect the criticalities.
Published experimental works show quite different
elementary excitation spectra, ranging from very small to very large
relaxation rates as compared to bulk excitation spectra
\cite{wu,schwalb}.
Up to now, relatively small number of theoretical studies
treating the  fluctuation dymamics in confined
liquid crystals near the bulk coexistence
have been published. Recently,
Ziherl and \u{Z}umer\cite{ziherl} have published a work dealing with
order-parameter pretransitional fluctuations
above $T_{NI}$ under the condition of
a strong homeotropic anchoring of the nematic director.
Apart from the typical soft fluctuation mode, reflecting the
interface position fluctuations,
their numerical analysis also indicated slow uniaxial
director modes located close to the substrate.
In the discussed work the critical wetting regime is
ensured by a strong scalar order-parameter anchoring at the surface,
describing substrates enhancing the nematic order.
Another extreme type of substrate anchoring
yet allowing for a critical wetting regime,
 called here
{\em weak-anchoring limit},
corresponds to substrates strongly suppressing the nematic
arrangement. For example, substrates
treated by the $S_iO$-evaporation
technique may produce such disorder effects.
In this case (and under some additional conditions discussed below),
it is the isotropic phase which wets
the wall, so that one may expect different behavior
of the normal fluctuation modes.

A goal of the present work is to compare the fluctuation mode
dynamics based on such extremely different
substrate potentials. The emphasis is especially on substrates
strongly suppressing the nematic order.
A characteristic feature of the  weak anchoring
is the possible critical behavior
of the surface scalar order parameter by itself.
This may also be of importance in the critical wetting experiments,
as the finite size effects frequently
mask the weak logarithmic
divergence in the interface position (characteristic for
short-ranged substrate potentials)
\cite{lipowsky2}.
As argued below, in the weak-anchoring limit,
when the isotropic phase wets
completely the substrate, in addition to the
scalar order-parameter soft excitation,
{\em gapped biaxial director modes}
located in the presurface quasi-isotropic layer can exist.
Basically, the low-energy spectrum of these local
biaxial excitations is determined
by the mean-field interface position. Since the latter
is logarithmically divergent at $T=T_{NI}$, the excitation
spectrum shows a characteristic cusplike behavior in that limit,
as found also numerically in the strong-achoring case \cite{ziherl}.

The cited results are based on the Gaussian approximation and
suggest constant nematic-director mean-field configurations,
when the director fluctuations decouple from the "dangerous"
scalar order-parameter modes.
In principle, a series of effects can invalidate
the above two approximations. At a first place,
higher-order fluctuation effects
may be of importance since $D=3$ is the upper critical spatial
dimension for a critical wetting transition.
In particular, it is well-known
that the effective nematic-isotropic interfacial width
diverges in $D=3$ \cite{brezin}.
Clearly, this latter divergence can modify
the local excitation spectra.
However, close enough to $T_{NI}$
the simple Gaussian picture of local director excitations
is still valid, since the effective interfacial width
is much smaller than the domain thickness (see below).
As to the inhomogeneous director
configurations, they can result from many sources,
the main effect being  the appearence of
intermode couplings. As an example,
we will discuss below order-electricity effects, as the problem
is characterized by strong scalar order-parameter variations
in the interfacial region.

The rest of the paper is organized as follows. In Section II
we define the surface Landau-de Gennes  free-energy functional and
the free-energy substrate potentials
describing the strong- and  weak-anchoring
limits. For uniform-director configurations,  we present
the mean-field scalar order-parameter
profiles which are later used to fix
the relevant areas of the phase diagrams.
In Section III we study the Gaussian
order-parameter fluctuations.
The emphasis is on the properties of the local nematic-director
excitations in the two cases of substrate anchorings and
in complete wetting regimes. The role of
interface position fluctuations and order electricity effects
are also analyzed here. The last Section contains discussions and
a summary of the results.

\section{Landau-de Gennes model}
We adopt the macroscopic Landau-de Gennes description (one
elastic-constant approximation) based on
the following free-energy density
 \cite{degennes}:
\begin{equation}
f_{LG}({\bf Q})=\frac{L}{2}(\partial_{\alpha}
Q_{\beta\gamma})^2+\frac{a}{2}
Q_{\alpha\beta}Q_{\alpha\beta}-\frac{b}{3}
Q_{\alpha\beta}Q_{\beta\gamma}Q_{\gamma\alpha}+\frac{c}{4}
(Q_{\alpha\beta}Q_{\alpha\beta})^2.
\label{lg}
\end{equation}
Here ${\bf Q}=Q_{\alpha\beta}({\bf r})$
denots the nematic order-parameter field (symmetric
traceless second-rank tensor),
$L$ is an elastic constant,
and $a\equiv \alpha (T-T^*)$. The material constants
$\alpha$, $b$,
and $c$ are supposed to be positive and temperature-independent.
$T^*$ is the supercooling temperature for the bulk isotropic phase.

It is convenient to introdice the following reduced quantities:
(i) the reduced tensor order parameter
${\bf S}\equiv {\bf Q}/Q_c$, where $Q_c=2b/3\sqrt{6}c$ is the
value of the scalar order parameter $Q$
at the bulk nematic-isotropic
phase-transition temperature $T_{NI}=T^*+b^2/27\alpha
c$, (ii) the reduced temperature $\tau=(T-T_{NI})/(T_{NI}-T^*)$.
The liquid crystal is assumed to occupy the semi-infinite
space $z>0$, $z$ being the coordinate normal to the surface.
Using the above reduced quantities, the surface
free-energy functional can be recasted to the
following form:
\begin{equation}
F_s[{\bf S}]=LQ_c^2
\int dxdy \int^{\infty}_0 dz\left\{ \frac{1}{2}(\partial_{\alpha}
S_{\beta\gamma})^2+\frac{1}{\xi_0^2}\left[ f({\bf S})-f({\bf S}_{b})
\right] +\delta(z)f_0({\bf S})\right\},
\label{fe}
\end{equation}
where the uniform  free-energy potential $f({\bf S})$ reads
\begin{equation}
  f({\bf S})=\frac{1+\tau}{2}
{\rm tr}\,{\bf S}^2-\sqrt{6}{\rm tr}\,{\bf S}^3
+\frac{1}{2}\left[ {\rm tr}\,{\bf S}^2\right]^2.
\label{bc}
\end{equation}
Here ${\bf S}_{b}={\bf S}_{b}(\tau)$ denotes
the order parameter deeply
in the bulk ($z\rightarrow \infty$). The correlation length of the
isotropic phase at coexistence is denoted by
$\xi_0=\sqrt{L/\alpha (T_{NI}-T^*)}$.
The short-ranged substrate potential $f_0({\bf S})$
(defined at $z=0$) comes from
a variation of the intermolecular
forces near the surface.
Up to second order in ${\bf S}$, $f_0({\bf S})$ is frequently
modeled by the expression \cite{nobili}
\begin{equation}
f_0({\bf S})=\frac{a_s}{2}{\rm tr}\,({\bf S}-{\bf S}_s)^2,
\label{f0}
\end{equation}
where ${\bf S}_s$ is a symmetric traceless second-rank
tensor characterizing the
substrate, and $a_s$ is a phenomenological constant
related to the ratio of the coupling
constants at the surface to those in the bulk.

The Euler-Lagrange equation
related to Eq. (\ref{fe}) reads
\begin{equation}
-\xi_0^2\Delta {\bf S}+
(1+\tau){\bf S}+3\sqrt{6}\,\left[{\bf S}^2-
\frac{1}{3}\left( {\rm tr}\, {\bf S}^2\right){\bf 1}\right]
+2\left({\rm tr}\, {\bf S}^2\right){\bf S}=0,
\label{el}
\end{equation}
where ${\bf 1}$ is the unit second-rank tensor. The boundary
condition at $z=0$, related to the last equation, is
\begin{equation}
\left[ -\frac{d{\bf S}}{dz}+\frac{\partial f_0({\bf S})}
{\partial {\bf S}}\right] \bigg|_{z=0}=0.
\label{bound}
\end{equation}


\subsection{Uniform-director order-parameter profiles}

In a uniaxial uniform-director
mean-field configuration,
the  solution of Eq. (\ref{el}) has the following
generic form:
\begin{equation}
{\bf S}_0(z)=\frac{u(z)}{\sqrt{6}}
(3{\bf n}_0\otimes {\bf n}_0-{\bf 1}),
\label{s0}
\end{equation}
where ${\bf n}_0$ is the  constant nematic director and $u(z)$
is the  scalar order-parameter profile.
 As discussed below,
the uniform-director ansatz (\ref{s0}) is
presumably not valid when Eq. (\ref{fe}) contains terms imposing
elastic nematic-director distorsions
(such as  order-electricity or
flexoelectric free-energy terms). Nevertheless, the uniform-director
approximation is a useful starting point even in the presence of
such terms, as far as the system
is in a regime of critical wetting and
close to  $T_{NI}$.

For future references, we briefly study below the
mean-field solutions of Eq. (\ref{el}) as well as  the
surface phase diagrams (for the two  extreme
cases of surface free-energy
potentials discussed above) in a vicinity of the bulk
transition temperature $T_{NI}$.
In a uniform-director approximation,  Eq. (\ref{el})
reduces to the following non-linear differential equation
for the profile function $u(z)$:
\begin{equation}
-\frac{d^2 u}{dz^2}+\frac{1}{\xi_0^2}\frac{\partial f(u)}{\partial
u}=0,
\label{eu}
\end{equation}
\begin{equation}
f(u)=\frac{1+\tau}{2}u^2-u^3+\frac{1}{2}u^4,
\label{f}
\end{equation}
supplemented by the  boundary condition
\begin{equation}
\frac{du}{dz}\bigg|_{z=0}=\frac{\partial f_0}
{\partial
u_0}.
\label{bu}
\end{equation}
Here  $u_0$ is the value of the scalar order parameter
$u(z)$ at the surface,
$u_0 \equiv u(z)_{z=0}$.
Deeply in the bulk  the  conditions
read $\left( du/dz\right)_{z=+\infty}=0$ and
$u(z=+\infty)=u_b$, where
$u_b=u_b(\tau)$ is the bulk scalar order parameter: $u_b=0$ in the
isotropic phase and
\begin{equation}
u_b(\tau)=\frac{1}{4}\left(3+\sqrt{1-8\tau }\right),
\hspace{1cm}\tau\le 0,
\end{equation}
in the ordered nematic phase.

It is instructive to rewrite Eq. (\ref{eu}), by use of
the boundary conditions for $u(z)$, in the form
\begin{equation}
\xi_0\frac{du}{dz}={\rm sign}(u_b-u_0)\sqrt{ 2f(u)-2f(u_b)},
\label{eu1}
\end{equation}
where ${\rm sign}(u_b-u_0)=\pm 1$ for $u_b>u_0$ and $u_b<u_0$,
respectively.
Now, using Eq. (\ref{eu1}),
the boundary condition (\ref{bu}) can be transformed to an
implicit equation for $u_0$,
\begin{equation}
{\rm sign}(u_b-u_0)\frac{1}{\xi_0}\sqrt{ 2f(u_0)-2f(u_b)}=
\frac{\partial f_0}{\partial u_0},
\label{bu1}
\end{equation}
and the free-energy functional (\ref{fe}) reduces to the
following function of $u_0$:
\begin{equation}
f_s(u_0)\equiv \frac{F_s[u_0]}{LQ_c^2}=\frac{{\rm sign}(u_b-u_0)}{\xi_0}
\int^{u_b}_{u_0} du\sqrt{
2f(u)-2f(u_b)}+f_0(u_0).
\label{fs}
\end{equation}
In terms of the function $f_s(u_0)$,
Eq. (\ref{bu1}) is equivalent to the minimum condition
$df_s(u_0)/du_0=0$,
whereas the phase stability condition reads
\begin{equation}
\frac{d^2f_s(u_0)}{du_0^2}\equiv \frac{
{\rm sign} (u_0-u_b)}{\xi_0
\sqrt{2f(u)-2f(u_b)}}\frac{df(u_0)}{du_0}+\frac{d^2f_0(u_0)}{du_0^2}>0.
\label{seq}
\end{equation}

The last three equations completely
describe the  surface phase diagram of a semi-infinite nematic
liquid crystal in uniform-director uniaxial states.
In the  context of surface phase transitions
in nematic liquid crystals,
using the substrate potential $f_0(u_0)=-h_0u_0$,
these  equations
were originally analyzed by Sheng \cite{sheng}.
Surface free-energy  potentials of the form
\begin{equation}
f_0(u_0)=-h_0u_0+\frac{a_s}{2}u_0^2,
\label{wp}
\end{equation}
constitute the basis of the theory of {\em wetting fenonomena}
in simple liquids\cite{pandit} and they have also been used
in the context of liquid crystals by Sluckin and Poniewierski
\cite{sluckin}. Since the emphasis in the present study is on
fluctuation effects in a regime of critical wetting,
we do not analyze the full phase diagram related to Eq. (\ref{f0}).
Instead, we mainly consider uniaxial
surface potentials which can be described by
the tensor parameter
\begin{equation}
{\bf S}_s=\frac{u_s}{\sqrt{6}}
(3{\bf n}_s\otimes {\bf n}_s-{\bf 1})
\label{ss}
\end{equation}
in Eq. (\ref{f0}),where $u_s$ and ${\bf n}_s$
are the scalar order parameter
and the director orientation preffered at the surface. To
be concrete, in what follows we assume that the unit vector ${\bf n}_s$
is normal to the surface (homeotropic director anchoring).
In the cases without elastic distorsion
forces, such as those described by the free-energy potential
(\ref{lg}) under appropriate boundary conditions,
the director  ${\bf n}_0$ is always parallel
to ${\bf n}_s$, provided that $u_s > 0$.
In this case Eq. (\ref{f0}) is reduced to the  following
substrate potential:
\begin{equation}
f_0(u_0)=\frac{a_s}{2}(u_0-u_s)^2,\hspace{0.5cm} a_s \ge 0.
\label{f01}
\end{equation}

There are two interesting  limits
of Eq. (\ref{f01}) when the  solution of Eq. (\ref{bu1})
is extremely simple: (i) The {\em srong-anchoring limit},
$a_s\rightarrow \infty$, which
is appropriate for substrates imposing
 a strong scalar order-parameter anchoring at the surface.
In this case the solution of Eq. (\ref{bu1}) just reads $u_0=u_s$.
This is also the case discussed in Ref. \onlinecite{ziherl}.
(ii) The second interesting limit of Eq. (\ref{f01})
corresponds to the choice $u_s=0$.
This latter condition may fit to, say,
substrates treated by $S_iO$-
evaporation technique, when  the nematic order is strongly
suppressed. In this case Eq. (\ref{f0})
reduces to
$f_0({\bf S})=(a_s/2){\rm tr}\,{\bf S}^2$, so that the substrate
does not impose any preferred direction on the nematic director.
We call this {\em weak-anchoring limit}.
It is important that
the above simplified potentials generate the main
critical wetting regimes\cite{lipowsky} found with
the more general form (\ref{wp}),
so that we restrict
our interest to the above extremely different cases.

For the bulk free-energy potential (\ref{f}) and for
uniform-director mean-field configurations,
the non-linear differential equation (\ref{eu1})
can be solved both above and below
$T_{NI}$ and for arbitrary $u_0$. The surface
scalar order parameter
$u_0$ should be independently obtained from Eqs.
(\ref{bu1}) and (\ref{f01}).
The explicit expression for
the  order-parameter profile above $T_{NI}$ reads
\begin{equation}
u(z)=\frac{1+\tau}{1+\sqrt{\tau}\,\sinh\left( z/\!\xi_I
+\alpha_I\right)},
\hspace{1cm} \tau>0,
\label{u+}
\end{equation}
\[
 \alpha_I={\rm arcsinh} \left[\frac{1}{\sqrt{\tau}}\,\left(
\frac{1+\tau}{u_0}-1\right)\right],
\]
where
$\xi_I=\xi_0/\sqrt{1+\tau}$ is the correlation length in the
isotropic phase.

Below the bulk coexistence temperature $T_{NI}$ the
profile function reads
\begin{equation}
u(z)=u_b-\frac{3u_b-2(1+\tau)}{2u_b-1-\sqrt{u_b-1}\,{\rm
sinh} \left[ {\rm sign}(u_0-u_b)z/\!\xi_N+\alpha_N\right]},
\hspace{1cm} \tau<0,
\label{u-}
\end{equation}
\[
 \alpha_N={\rm arcsinh} \left\{\frac{1}{\sqrt{u_b-1}}\left[
2u_b-1+\frac{3u_b-2(1+\tau)}{u_0-u_b}\right]\right\},
\]
where
$\xi_N\equiv d^2f(u_b)\!/du_b^2=\xi_0/\!\sqrt{3u_b-2(1+\tau )}$
is the correlation length in the
nematic phase.
For $\tau>0$ the profile $u(z)$ describes a
paranematic layer, whereas the bulk
is occupied by the isotropic phase.
For large enough $u_0$, $u(z)$ has an inflection
point at $z=d_I$ marking the center of the nematic-isotropic
interface. The implicit equation for $d_I$
is $u(d_I)=u_1$, where the function
$u_1=u_1(\tau)=(3-\sqrt{1-8\tau })/4$ corresponds to the
local maximum of $f(u)$, Eq. (\ref{f}). Explicitly,
\begin{equation}
d_I=\xi_I\ln \left| \left(\frac{u_0}{u_1}\right)
\frac{\tau+(1-u_1)+\sqrt{
(1+\tau)\left[\tau+(1-u_1)^2\right]}}{\tau+(1-u_0)+\sqrt{
(1+\tau)\left[\tau+(1-u_0)^2\right]}}\right|,
\hspace{0.5cm} u_0>u_1.
\label{li}
\end{equation}

Similar formulae can be established below $T_{NI}$. For $u_0<u_b$
the surface layer is less ordered as compared to the bulk.
If in addition $u_0<u_1$, the profile has an
inflection point at $z=d_N$ marking the interface position
difined by the equation
$u(d_N)=u_1$ (see Fig. 1). The result reads
\begin{equation}
d_N=\xi_N \ln \left| \left(\frac{u_b-u_0}{u_b-u_1}\right)
\frac{u_1u_b+(u_b-1)(2u_b+u_1)+\sqrt{u_b(4u_b-3)\left[
u_b-1+(u_b+u_1-1)^2 \right]}}{u_0u_b+(u_b-1)(2u_b+u_0)+
\sqrt{u_b(4u_b-3)\left[
u_b-1+(u_b+u_0-1)^2\right]}}\right|.
\label{ln}
\end{equation}

\subsection{Surface phase diagrams near $T_{NI}$}

The profile functions, Eqs. (\ref{u+}) and (\ref{u-}),
may be used to describe the surface phase diagrams.
In the context of
interface delocalization transitions \cite{note1},
the surface phase diagram has already been analyzed in the
framework of the  substrate potential (\ref{wp}) \cite{lipowsky}.
For our study it is enough to consider the extreme
cases specified above and to mark the respective peculiarities
connected to the form of the used surface free-energy potentials.

(i) {\em Strong-anchoring limit}:\\
Firstly, one must find
the surface order parameter $u_0$ from Eq. (\ref{bu1}).
In the limit $a_s\rightarrow+\infty$ the solution of
Eq. (\ref{bu1}) just reads $u_0=u_s$,
so that the surface is characterized by the parameter $u_s$.
Let us consider the phase diagram for $\tau>0$.
If $u_s> 1$ and $\tau \rightarrow 0^+$,
Eq. (\ref{li}) predicts a logarithmically diverging
interface position  given by $d_I= \xi_0\ln(1\!/\tau)+O(1)$.
Exactly at $u_s=1$  one finds
$d_I=(\xi_0/2)\ln (1/\!\tau)+O(1)$. The
critical behavior is connected to
the following non-analytic terms in the surface
free energy (\ref{fs}) considered as a function of
$\tau$ and $\Delta \equiv u_s-1$:
\begin{equation}
f_s^{sing}(\tau, \Delta)=
\frac{\tau}{2}\ln \frac{\Delta+\sqrt{\Delta^2+\tau}}{\tau}+
\frac{\Delta}{2}\sqrt{\Delta^2+\tau}+\frac{1}{3}(\Delta^2+\tau)^{3/2}.
\label{fss}
\end{equation}
In the region $\Delta<0$  there are no singular terms at all,
and the thickness of the presurface nematic layer $d_I$
is of the order of the correlation length $\xi_I$ (partial wetting).
On the other hand, one finds $f_s^{sing}\propto \tau\ln (1/\tau)$
as  $\Delta \ge 0$. Thus, there is
a line of critical points,
defined by $\tau =0^+$ and $u_s\ge 1$, describing a critical wetting
of the substrate by the nematic phase.
The transition at $\tau =0^+$ for $u_s> 1$ is  known as
a {\em complete wetting}
or as a {\em critical interface delocalization
transition} of type $P^+$. \cite{lipowsky}
The point $(\tau ,\Delta)=(0^+,0)$
is a special point on the phase diagram since
the parameter $\Delta$ becomes  a
 relevant field (in addition to $\tau$): the derivative $\partial^2
f_s/\partial \Delta^2 \propto \Delta/
\sqrt{\tau+\Delta^2}$ has a finite
jump at $\Delta =0$.
Respectively, the interface position
$d_I =\xi_0\ln (1/|\Delta|)+O(1)$ is finite as $\Delta<0$,
whereas it is delocalazed
(infinite) when $\Delta >0$.
As a matter of fact, the multicritical point
$(\tau,\Delta)=(0^+,0)$ corresponds to
the tricritical point $T^+$ found for the substrate
potential (\ref{wp}), but in the strong-anchoring limit ($a_s\rightarrow
\infty$) the summetry breaking field $h_0$ is irrelevant.

Similar critical properties can be established just
below the bulk coexistence temperature $T_{NI}$.
Now the interface is located at $z=d_N$,
and the region $0<z<d_N$ is occupied by a suppressed (as compared to the
bulk)
nematic phase, provided that $u_s< u_1$. For positive
values of the parameter $u_s$, $d_N$ remains finite at $\tau=0^-$,
and the thickness of the surface layer $d_N$ is of the order of
$\xi_N$.
Complete wetting of the substrate by the isotropic phase
is possible only  for non-positive
values of $u_s$\cite{note2}. The point
$(\tau,u_s)=(0^-,0)$ is a multicritical point analogous to those
discussed above. The interface position
$d_N$ shows the same asymptotic behavior as $d_I$ does:
for $u_s<0$ and $\tau \rightarrow 0^-$,
it diverges as $d_N= \xi_0\ln(1/\!|\tau |)+O(1)$,
whereas at $u_s=0$ one has
$d_N=(\xi_0/2)\ln(1/\!|\tau |)+O(1)$. For fixed $\tau =0^-$,
$d_N$ is finite when $u_s>0$,  $d_N= \xi_0 \ln (1/u_s)+O(u_s)$, and
it becomes infinite as $u_s<0$.

(ii) {\em Weak-anchoring limit}:

This case is more interesting in a sense that now
the surface order parameter $u_0$ is not pinned by the
surface and may change with the temperature, $u_0=u_0(\tau)$.
As obtained from the inplicit Eq. (\ref{bu1}),
the function $u_0(\tau)$
has the following asymptotic behavior. For large enough
$a_s$ ($a_s>\xi_0^{-1}$) and $\tau \rightarrow 0^-$, $u_0=u_0(\tau)$
vanishes as
\begin{equation}
u_0(\tau)=\frac{1}{\sqrt{\xi_0^2a_s^2-1}}\left|\tau \right|^{\frac{1}{2}}
+O(\left|\tau\right|),\hspace{0.5cm} a_s\xi_0>1,\hspace{0.5cm}
\tau<0,
\label{u01}
\end{equation}
whereas $u_0=0$ all over the region $\tau >0$.
This additional surface  criticality is specific for the
weak-anchoring limit and
takes place simultaneously with the critical interface
delocalization discussed above.
The asymptotic behavior of $d_N$ as $\tau \rightarrow 0^-$ (see Fig. 1)
is analogous to those found
in the case of srong anchoring:
$d_N= \xi_0\ln(1/\!u_0)+O(1)$, with $u_0=u_0(\tau)$ from
Eq. (\ref{u01}).

The case $a_s <\xi_0^{-1}$ is less interesting in a sense that now
the surface  order parameter $u_0$
is discontinuous at $T_{NI}$: it jumps from the finite
value $u_0=1-a_s\xi_0+O(|\tau|)$ (for
$\tau \rightarrow 0^-$) to zero (for $\tau >0$). Respectively,
there is no  critical behavior for $a_s\xi_0<1$,
and $d_N$ is of the order of the correlation length
in the nematic phase $\xi_N$. The point
$(\tau, a_s\xi_0)=(0^-,1)$
 is a milticritical  point (type $T^-$ according
 to Ref. (\onlinecite{lipowsky})) characterized by three relevant
 fields, i.e., $\tau$, $a_s\xi_0$, and $h_0$.
Exactly at $a_s\xi_0=1$,
the surface order parameter $u_0(\tau)$
acquires the asymptotic behavior
\begin{equation}
u_0(\tau)=\left|\frac{\tau}{2}\right|^{\frac{1}{3}}
+O(\left|\tau\right|),\hspace{0.5cm} a_s\xi_0=1,\hspace{0.5cm}
\tau<0.
\label{u02}
\end{equation}
Respectively, the interface position delocalizes as
$d_N= \xi_0\ln(1/\!u_0)+O(1)$), with $u_0=u_0(\tau)$ defined by
Eq. (\ref{u02}).

\section{Local fluctuation modes}
Within the Landau-de Gennes approach the normal fluctuation modes
are obtained by solving the eigenvalue problem connected to
the linearized Euler-Lagrange
equation. The latter equation can be derived from  Eq. (\ref{el})
by use of the
order-parameter decomposition
\begin{equation}
{\bf S}({\bf r}) ={\bf S}_0(z)+\mbox{\boldmath $\phi$ } ({\bf r}).
\label{lin}
\end{equation}
Here ${\bf S_0}(z)$ is the mean-field solution (\ref{s0}),
as defined by the scalar order parameter $u(z)$,
Eqs. (\ref{u+}) and (\ref{u-}). $\mbox{\boldmath $\phi$ }
({\bf r})$ is the order-parameter
fluctuation field (symmetric traceless
second-rank tensor).

It is useful to adopt the following parametrization of
$\mbox{\boldmath $\phi$} ({\bf r})$\cite{pokrovskii}:
\begin{equation}
\mbox{\boldmath $\phi$}
({\bf r})=\sum_{i=0}^4 \phi_{i}({\bf r}){\bf g}_i,
\end{equation}
where the base tensors ${\bf g}_i$, $i=0,\cdots , 4$ read
\[
{\bf g}_0=\frac{1}{\sqrt{6}}(3{\bf n}_0
\otimes {\bf n}_0-{\bf 1}),
\]
\begin{equation}
{\bf g}_1=\frac{1}{\sqrt{2}}({\bf e}_1
\otimes {\bf n}_0+{\bf n}_0\otimes
{\bf e}_1),\hspace{0.5cm}
{\bf g}_2=\frac{1}{\sqrt{2}}({\bf e}_2
\otimes {\bf n}_0+{\bf n}_0\otimes
{\bf e}_2),
\end{equation}
\[
{\bf g}_3=\frac{1}{\sqrt{2}}({\bf e}_1\otimes
{\bf e}_1-{\bf e}_2\otimes
{\bf e}_2),\hspace{0.5cm}
{\bf g}_4=\frac{1}{\sqrt{2}}({\bf e}_1\otimes
{\bf e}_2+{\bf e}_2\otimes
{\bf e}_1),
\]
and satisfy the orthogonality
relation ${\rm Tr}\,({\bf g}_i.
{\bf g}_j)=\delta_{ij}$.
The unit vectors ${\bf e}_1 \bot {\bf e}_2$
are perpendicular to the
nematic director ${ \bf n}_0$.

The variable $\phi_0({\bf r})$ describes
{\em longitudinal}
scalar order-parameter fluctuations,
whereas the pair variables
$[\phi_1({\bf r}),\phi_2({\bf r})]$
and $[\phi_3({\bf r}),\phi_4({\bf r})]$
are connected to the {\em transverse}
uniaxial and biaxial director
fluctuations, respectively.
Due to the symmetry in the $(x,y)$ plane, it is
suitable to work in a mixed
$({\bf k}_{\bot}, z)$ representation, i.e.,
 $\phi_i=\phi_i({\bf k_{\bot}},z)$,
where ${\bf k}_{\bot}=(k_x,k_y)$
 is the wave vector in the
$(x,y)$ plane. Now, projecting  the
linearized Euler-Lagrange equation
on the base tensors ${\bf g}_i$,  one obtains the
following independent Schr\"odinger-type eigenmode equations:
\begin{equation}
\left[ -\xi_0^2\frac{d^2}{dz^2}+
V_i(z)\right]\phi_i(z)
= E_i\phi_i(z), \hspace{0.5cm} i=0,\cdots ,4,
\label{ne}
\end{equation}
where the functions  $V_0(z)=1+\tau-6u(z)+6u(z)^2$,
$V_1(z)=V_2(z)=1+\tau-3u(z)+2u(z)^2$, and
$V_3(z)=V_4(z)=1+\tau+6u(z)+2u(z)^2$ may be
thought of as potential energies in a
related quantum-mechanical problem.
The profile function $u(z)$ is defined by
Eqs. (\ref{u+}) and (\ref{u-}).
The correlation functions  of the order-parameter fields
$G_i(k_{\bot};z,z^{\prime})=\left< \phi_i({\bf k}_{\bot},z)
\phi_i({\bf k}_{\bot},z^{\prime})\right>$ satisfy the equations
\begin{equation}
\left[k_{\bot}^2 -\xi_0^2\frac{d^2}{dz^2}+
V_i(z)\right]G_i(k_{\bot};z,z^{\prime})
= \delta (z-z^{\prime}), \hspace{0.5cm} i=0,\cdots ,4,
\label{gf}
\end{equation}
and can be represented, using the eigenmodes  $\phi_i^{(n)}(z)$
and the eigenvalues $E_i^{(n)}$ of Eq. (\ref{ne}), in the form
\begin{equation}
G_i(k_{\bot};z,z^{\prime})=
\sum_n \frac{\phi_i^{(n)*}(z)\phi_i^{(n)}
(z^{\prime})}{k_{\bot}^2+E_i^{(n)}},
\hspace{0.5cm} i=0,\cdots ,4.
\end{equation}
In what follows we study the eigenvalue problem
in the two extreme cases of surface free-energy potentials
specified above. In particular, we argue
that the local uniaxial director excitations are characteristic
of surface potentials favoring the nematic state,
whereas in the case of weak-anchoring potentials suppressing the
nematic order, as a rule, the local biaxial exciations
are favored.
In principle, the mode spectrum may be disturbed
by several issues, such as
higher-order fluctuation
effects, finite-size corrections, long-range substrate forces,
and the order electricity. Some
of the mentioned effects modifying the local director excitations
are discussed below. As to the external electromagnetic fileds,
this problem deserves a separate study.

\subsection{Strong-anchoring limit: local uniaxial director modes}

As discussed above, for substrates enhancing the scalar
order parameter ($u_0=u_s>1$) (and for a fixed surface
homeotropic orientation of ${\bf n_0}$),
the surface phase diagram contains a line of critical
points describing a complete wetting. The related
eigenvalue  problem, Eq. (\ref{ne}),  has recently
been numerically analyzed  by Ziherl and \u{Z}umer\cite{ziherl}.
In this strong-anchoring case,
the fluctuation modes are pinned at the surface, i.e.,
$\phi_i(z)|_{z=0}=0$, $i=1,\cdots,4$.
Two types of local exitations, related to the potential wells shown
in Fig. 2,
have been established. The
first one is the lowest soft mode $\phi_0^{(0)}(z)$
characterized by an energy $E_0^{(0)}(\tau)
\propto \tau$ and  related to the scalar
order-parameter field: This
mode describes fluctuations of the mean-field
interface position located at $z=d_I$. Physically,
this excitation appears as a result of the broken translational
symmetry [ $u(z)\not= 0$] and its existence is a typical
feature of the critical wetting transition.
The second type
of localized modes found in the discussed work are
connected with the uniaxial director fluctuation fields
$\phi_1(z)$ and $\phi_2(z)$. In the vicinity of $T_{NI}$ the
uniaxial modes soften and  become
gapless at  $T_{NI}$.
As seen below, these features of the local uniaxial director excitations
can easily be obtained analytically in the limit $\tau \rightarrow 0^+$.

The physics behind the local uniaxial director modes
is simple. At a mean-field level and in the limit
$\tau \rightarrow 0^+$,
the interface width, as a rule being of the order of
the correlation length $\xi_0$,
 is much smaller than the
layer thickness $d_I \simeq \xi_0\ln (1/\tau)$,
so that the
region $0<z<d_I$ may be thought of as a nematic plate of
width $d_I$, Eq. (\ref{li}). In the same limit, and for
the low-energy excitations, one can use
the following simpified form of the
potential $V_1(z)$ (see Fig. 2):
$V_1(z)\approx 0$ in the interval
$0<z<d_I$, and $V_1(z)\approx 1$ for $z>d_I$.
Thus, using the continuity of the logarithmic derivative of
the field $\phi_1(z)$ at $z=d_I$,
one finds the excitation energies $E^{(n)}_{1}=\xi_0^2k_n^2$.
The parameter $k_n$ satisfies the implicit equation
\begin{equation}
k_n d_I=\pi n-{\rm arcsin}(k_n\xi_0), \hspace{0.5cm}
n=1,2,\cdots, n_{max},
\label{kn1}
\end{equation}
where the number of localized modes
$n_{max}$ is finite, fixed by the
condition $0\le k_n\xi_0\le 1$, and
dependent on the reduced temperature
$\tau$ ( since $d_I=d_I(\tau)$).
In the limit $\tau \rightarrow 0^+$, when
$d_I\simeq \xi_0\ln (1/\tau)$,  the excitation energies $E^{(n)}_{1}$
take the followind asymptotic form:
\begin{equation}
E_1^{(n)}\simeq \frac{\pi^2n^2}{\ln^2 (1/\tau)},
\hspace{0.5cm}n=1,2,\cdots, n_{max}.
\label{e12}
\end{equation}
This expression is in agreement with the numerical
results of Ref. \onlinecite{ziherl} and  reproduces, in particular,
the observed cusplike bihavior of
the low-laying energy levels. We see that the low-energy
levels give a direct information about
the logarithmically divergent interface position,
$d_I\simeq \xi_0\ln (1/\tau)$. This
critical behavior is characteristic for the short-ranged substrate
interactions. For example, the Van der Waals type interactions
imply the power-low critical behavior $d_I\propto \tau^{-1/3}$, whereas
the Fr\'eedericksz-type wetting transition was shown
to give the asymptotic form $d_I\propto\tau^{-1/2}$ \cite{sluckin2}.
It is clear that the above excitation spectra
can  give  a valuable information on  both
the nature of presuface forces and the criticality by itself.
\subsection{Weak-anchoring case: local biaxial director modes}

Let us remember that in the case of
weak-anchoring substrate potentials
the complete wetting regime is realized  for
$a_s\ge \xi_0^{-1}$ as  $\tau \rightarrow
0^-$. Now it is the isotropic phase which wets the
wall and the presurface layer of thickness $d_N=d_N(\tau)$
may be thought of as a plate occupied by the isotropic phase.
As in the limit of strong anchoring,
the critical behavior is controlled by the lowest
scalar order-parameter mode $\phi_0^{(0)}$ describing
fluctuations of the mean-field interface position
at $z=d_N$. A good approximation for $\phi_0^{(0)}$
can be obtained in the limit $\tau\rightarrow 0^-$,
when the interface profile $u(z)$, Eq. (\ref{u-}),
practically coincides with the profile of the infinite system
at coexistence, $u(z)\approx u^{\infty}(z)=
\{1+\tanh[(z-d_N)/2\xi_0]\}$ (exlcuding a small
vicinity of the substrate). Since in the infinite system
the function $\phi_0^{\infty}(z)=du^{\infty}(z)/dz$ is a solution
of Eq. (\ref{ne}) with the potential $V_0(z)=V_0[u^{\infty}(z)]$,
it is clear that the variational ansatz
\begin{equation}
\phi_0^{(0)}(z)=C(\tau)e^{-z/\xi_0}+\frac{1}{4\cosh^2
[(z-d_N)/2\xi_0]},
\label{var}
\end{equation}
gives a good approximation for $\phi_0^{(0)}(z)$
in the limit $\tau\rightarrow 0^-$.
Here the  interface position  $d_N=d_N(\tau)$ is defined by Eq. (\ref{ln}).
The first term in Eq. (\ref{var}) is introduced in order to satisfy
the boundary condition (\ref{bu}), whereas the second one
is a solution of Eq. (\ref{ne}) for the infinite
system at $T=T_{NI}$.
From Eq. (\ref{bu}) one finds that the parameter $C=C(\tau)$
takes in the limit $ \tau \rightarrow 0^-$
the asymptotic form
 $C(\tau )=-(a_s\xi_0-1)u_0(\tau )$, where $u_0(\tau )\propto |\tau
 |^{1/2}$ is given by
Eq. (\ref{u01}) in the case $a_s\xi_0>1$.
Using the above ansatz as a variational function, one can
show analytically\cite{lipowsky4} that the lowest excited
state $\phi_0^{(0)}(z)$
has the energy $E_0^{(0)}\propto |\tau |$ and is connected
to fluctuations of the mean-field interface position at $z=d_N$.
More details on the importance of the discussed  soft
excitation will be presented in the next Subsection.

Now, let us address the nematic-director fluctuations.
Due to the fact that biaxial director fluctuations
are {\em strongly suppressed} in the bulk nematic phase
(as compared to the isotropic bulk phase),
there is a  well in the potential function $V_3(z)$ (see
Fig. 3) in the eigenmode equations (\ref{ne}).
The potential  well is located
in the presurface quasi-isotropic domain
with a characteristic thickness $d_N$.
Thus, in the weak-anchoring case
{\em bound biaxial director
modes} located in the persurface domain can appear.
On the other hand,
the uniaxial director modes, being gapless Goldstone modes
in the bulk nematic phase, are controlled by
the monotonically decreasing potential $V_1(z)$, so that
they will be delocalized all over the sample.
In the vicinity of $T_{NI}$ the low-energy levels of the
local biaxial modes  can easily be obtained
analytically by use of the same procedue applied above to
the uniaxial excitations in the regime of strong anchoring. Now the
potential $V_3(z)$ is simplified as $V_3(z)\approx 1$ (for $0<z<d_N$), and
$V_3(z)\approx 9$  (for $z>d_N$). The boundary condition at $z=0$ for the
biaxial field $\phi_{3}(z)$, as obtained from Eq. (\ref{bound}), reads
\begin{equation}
\left[ \frac{d\phi_{3}(z)}{dz}-a_s\phi_{3}(z)\right]_{z=0}=0.
\end{equation}
Using the same arguments as in the previous case,
one finds the excitation energies $E^{(n)}_{3}=1+\xi_0^2k_n^2$,
where the parameter $k_n$ now satisfies the followig implicit equation:
\begin{equation}
k_n d_N=\pi n-{\rm arcsin}\left(\frac{\xi_0 k_n}{2\sqrt{2}}\right)
-{\rm arcsin}\left(\frac{k_n}{\sqrt{k_n^2+a_s^2}}\right), \hspace{0.5cm}
n=1,2,\cdots, n_{max}.
\end{equation}
The number of localized modes $n_{max}$ is  fixed by the
condition $0\le k_n\xi_0/2\sqrt{2}\le 1$, and  $n_{max}$ also
depends on the reduced temperature
$\tau$ through the function $d_N=d_N(\tau)$.
Using the asymptotic expression
$d_N\simeq \xi_0\ln (1/|\tau |)$, one gets the following
result
in the limit $\tau \rightarrow 0^-$:
\begin{equation}
E_3^{(n)}=1+\frac{\pi^2n^2}{\ln^2 (1/|\tau |)},
\hspace{0.5cm}n=1,2,\cdots, n_{max}.
\label{e34}
\end{equation}
We see that in the limit $\tau \rightarrow 0^-$
the local biaxial  modes are stongly softened
as compared to the bulk biaxial fluctuations
in the nematic phase: For every $n$, the energies
$E_3^{(n)}$ change from $E_3^{(n)}=9$ (the gap in the bulk nematic
phase) to
$E_3^{(n)}=1$ (the gap in the bulk isotropic
phase). As expected, the local biaxial  mode
spectrum is gapped at $T=T_{NI}$.
\subsection{Role of the interface-position fluctuations}
 Since the upper  critical spatial dimension  for
a critical wetting transition is $D=3$\cite{lipowsky3,brezin},
higher order fluctuation
effects are not excluded. Indeed, as discussed above,
the singular part of the
mean-field free energy in the regime of complete wetting
behaves like $f_s^{sing}\propto -\tau \ln (\tau)$, Eq. (\ref{fss}).
On the other hand, the one-loop fluctuation contribution
to the  free energy
is related to the lowest soft mode
$\phi_0^{(0)}(z)$ since the energies of
the excited states of Eq. (\ref{ne})
are separated from  $E_0^{(0)}$ with a finite gap.
Using this fact, it is easy to see that
the fluctuation part of the free energy
is $f_s^{fl}(\tau )\propto
-E_0^{(0)}\ln (E_0^{(0)})\sim -\tau \ln (\tau)$,
so that it compares to the
singular part of the mean-field free energy.
Remember that for uniform-director
configurations
the scalar order-parameter field
$\phi_0({\bf r})$ is decoupled from the director fields
in the Gaussian approximation.
Therefore, in this approximation most of the results known
from the scalar theory\cite{lipowsky4}
can be  applied to nematic systems without changes.
However, the fluctuation mode  $\phi_0^{(0)}$
will effectively disturbe the studied
local director excitations.
It is easy to see this qualitatively if remember some of
the fluctuation effects connected to the
soft mode $\phi_0^{(0)}(z)$, originally obtained
in a context of the scalar order-parameter theory.  At the first place,
it can be shown that this mode
produces the singularities in the  correation
function
\begin{equation}
\left< \phi_0({\bf r}_{\bot},z_1)
\phi_0({\bf 0},z_2)\right>\sim \exp
(-r_{\bot}/\xi_{\|})\phi_0^{(0)}(z_1)\phi_0^{(0)}(z_2),
\end{equation}
where
$\xi_{\|}=1/\sqrt{E_0^{(0)}}\propto 1/\sqrt{\tau}$
 is the characteristic length of
the  scalar order-parameter correlations
parallel to the surface. Thus, there are critical long-range correlations
of the scalar order-parameter fluctuations parallel to the surface
(capillary waves). Another criticality connected to $\phi_0^{(0)}(z)$ is
the predicted divergence of the interfacial width.
Denoting by $d_I({\bf r}_{\bot})=
d_I+\zeta({\bf r}_{\bot})$ the local interface position, it can be
shown that the characteristic interface thickness
$\xi_{\bot}\equiv
\sqrt{<\zeta^2>}$ diverges as $\tau \rightarrow 0^+$
according the asymptotic form
$\xi_{\bot} \propto \ln \xi_{\|}\propto \sqrt{d_I}$\cite{brezin}.
Thise critical increase of the effective interface width, clearly,
will effectively change the potential $V_1(z)$ in Eq.
(\ref{ne}) which controls the  director excitation spectrum.
On the other hand, the threatment of the local fluctuation modes
used throughout the paper remains valid, since
close enough to $T_{NI}$ the effective interfacial width
will be much smaller than the characteristic domain
thickness.
\subsection{Role of the order electricity}
There is a series of issues which might invalidate the
uniform-director
appoximation suggested in the above analysis\cite{degennes1,skacej}.
The order electricity \cite{prost,barbero} belongs to this series.
Since the nematic-isotropic interface is characterized by a
strong variation of the scalar order parameter, one can expect that
the order-electricity effect will play an important role in the
wetting phenomena. Let us concentrate on the strong-anchorig limit,
when the nematic director is strongly anchored
 along the normal to the surface.
 The inhomogeneity  in the profile
function $u(z)$, Eq. (\ref{u+}), generates along the axis $z$
the electric polarization field $P^z_o=
\left( r_1{n_0}_z^2+r_2\right)(du/dz)$,
where $r_1$ and $r_2$ are the order-electric
coefficients \cite{barbero}.
Denoting by $\theta $ the polar angle of the director
${\bf n_0}=(\sin \theta, 0, \cos \theta)$, it is easy to see that
$\theta =\theta (z)$ since the free-energy density term $f_o$,
connected with $P^z_o$, mixes ${\bf n_0}$ and $du/dz$:
\begin{equation}
f_o=-\frac{1}{2} P^z_o E^z_o=\frac{ 2 \pi }{\epsilon_{zz}(\theta)}
\left( r_1\cos^2\theta+r_2\right)^2\left(\frac{du}{dz}\right)^2.
\end{equation}
Here $\epsilon_{zz}(\theta)=\epsilon_{\bot}
+(\epsilon_{\|}-\epsilon_{\bot})\cos^2\theta$ is the
$zz$ component of the dielectric tensor and $E^z_o$
is the $z$ component of the induced electic field. Combaining the
last equation with the elastic term in Eq. (\ref{fe}), one finds
the result that the
elastic constant $L$ in Eq. (\ref{fe}) is effectively
renormalized as $L\rightarrow
L_{eff}(\theta)= L+ \left[4\pi/\epsilon_{zz}(\theta)\right]
\left( r_1\cos^2\theta+r_2\right)^2$.
To find the function $\theta (z)$ we  will follow
the arguments of Ref. \onlinecite{barbero},
applicable in our case for $\tau \rightarrow 0^+$.
The Euler-Lagrange
equation for the polar angle
$\theta$, as obtained from  Eq. (\ref{fe}), reads
\begin{equation}
\frac{d^2\theta}{dz^2}=\frac{1}{6Lu^2}\frac{dL_{eff}}{d\theta}
\left(\frac{du}{dz}\right)^2.
\end{equation}
As shown above,  $du/dz\propto \phi_0^{\infty}(z)$
is a $\delta$-like function
centered at $z=d_I$, Eq. (\ref{var}).
Thus, the function $\theta (z)$ can be approximated with
a linear function in the region $0< z < d_I$, excluding
the vicinity of the interface and, possibly, of the surface.
Since the interfacial energy is proportional to $L_{eff}^{1/2}$,
the minimum condition for this energy
$dL_{eff}/d\theta = 0$  also defines
the average  polar angle at the interface $\theta_0$\cite{note3}.
We do not consider here possible deviations from
the linear behavior of $\theta(z)$ close to the substrate \cite{barbero1}.
Therefore, as $\tau \rightarrow 0^+$ the inhomogeneous director
state is characterized by the polar angle
\begin{equation}
\theta(z) =\theta_0\frac{z}{d_I}\simeq
\theta_0\,\frac{z}{\xi_0}\,\frac{1}
{\ln (1/\tau)}, \hspace{0.3cm}
\tau \rightarrow 0^+,\hspace{0.3cm} z\le d_I.
\end{equation}

The inhomogeneous director state described above can generate
several changes in the fluctuation mode dynamics. Since the
local base tensors are now coordinate-dependent, ${\bf g}_i={\bf g}_i(z)$,
the normal mode equations (\ref{ne}) are coupled. In particular,
there is a coupling between the longitudinal
scalar order-parameter field $\phi_0(z)$
and the transverse director fluctuations $\phi_2(z)$ and $\phi_3(z)$.
The mode coupling is, however, asymptotically small,
as it is controlled by the small
parameter $\xi_0/d_I\propto \ln^{-1} (1/\tau)$.
As the interaction decreases logarithmically ,
the effect of order-electicity
mode coupling can play an important role in  real experiments.
\section{Summary of the results and discussion}
In this article the emphasis was on specific features
of the presurface fluctuation mode dynamics in nematic
liquid crystals, arising in a
regime of critical wetting when the system is close to the
nematic-isotropic phase transition temperature.
The importance of substratate anchoring for the
director mode dynamics is demonstrated by use of
two extreme types of surface free-energy
potentials.

Our key results are as  follows:
In the case of uniform-director mean-field
configurations and for a homeotropic director anchoring,
surface free-energy potentials
strongly suppressing the orientational
order favor the appearence
of bound biaxial nematic-director
fluctuation modes located in the domain occupied
by the wetting thermodynamic phase. Instead,
substrate forces which enhance the orientational
order promote bound uniaxial modes in the presurface
domain. Close to the phase
coexistence temperature both types of local
director excitations are strongly softened (as compared to their bulk
counterparts)  and acquire
characteristic cusplike
low-energy spectra which are directly connected
to the critical behavior of
the mean-field interface position. Since the physics of
these local excitations closely reflects the wetting critical behavior,
they can be used as a tool
to examine the critical wetting phenomena
and the presurface interactions
in nematic liquid crystals.

We have considered one of the possible reasons for the experimentally
observed different spectra of elementary exitations,
ranging from very small to very large (as compared to the bulk)
relaxation rates. Clearly, there could be a number of other factors
modifying substantially the director exitation spectra.
We have already briefly discussed two of them, i.e., the higher-order
fluctuation effects and order electricity effects. As to
the interface position fluctuations, in the
upper spatial critical dimension $D=3$ they lead
to a divergence of the effective interface width. Thus,
the local excitation specta can be disturbed, as the potentials
$V_i(z)$ will be effectively changed as well.
In addition,
the fluctuation effects in $D=3$ are known \cite{brezin}
to depend on the value of the dimensionless parameter
$\omega =T/4\pi \xi \sigma $, where $\xi$ is the bulk
correlation length of the phase attracted to the wall and
$\sigma$ is the surface tension of the free interface.
It will be interesting to see the role of this parameter
in different liquid crystal materials.
As to the order-electricity effects, they are expected to play an
important role in the real experiments  due to
the strong variation of the scalar order parameter
in the interfacial region.
The coupling of the normal Gaussian modes
is the main effect of the inhomogeneity.
It is important that in this case the director fluctuation
modes are coupled with the "dangerous"
scalar order-parameter soft mode,
so that the picture of the Gaussian fluctuation dynamics
can be importantly disturbed.

In conclusion, the finite-size effects, the long-ranged substrate
interactions, and the external electromagnetic fields
belong to the class of factors that will modify
the discussed local fluctuation mode dynamics. Clearly,
these issues deserve future studies.
\acknowledgements
The author thanks M. P. Petrov and I. Dozov
for the numerous useful discussions. This study was supported
by the COPERNICUS Program (Grant No. ERBIC 15 CT 960744)
and the Bulgarian Science Foundation (Grant No. 582).

\pagebreak
\begin{figure}
\epsfxsize=10cm
\epsfbox{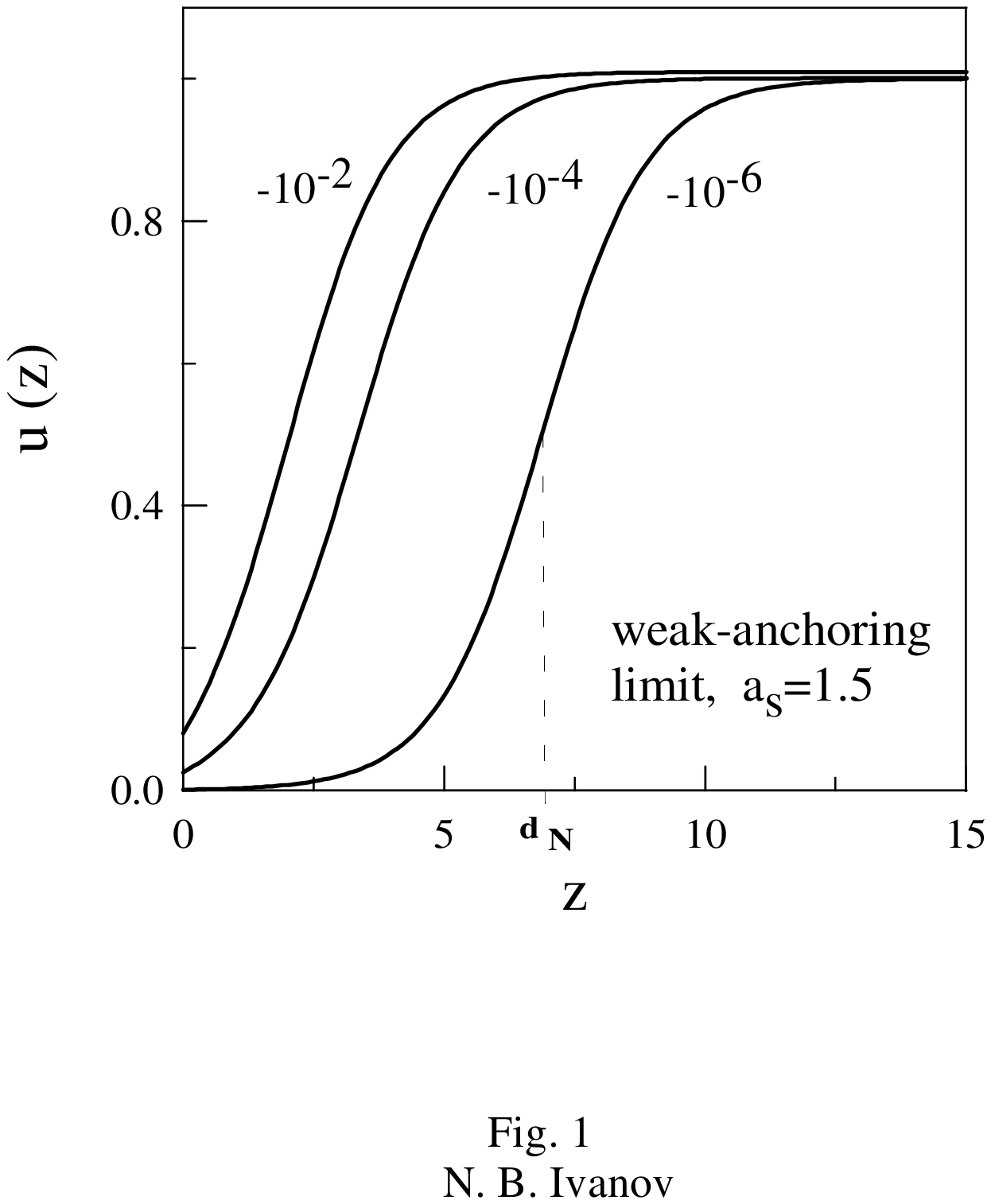}
\caption{Scalar order-parameter profiles for
the reduced temperatures $\tau =-10^{-2}$, $-10^{-4}$, and $-10^{-6}$
in the case of substrates strongly suppressing the
nematic order. The inflaction point at $z=d_{N}$ marks
the characteristic isotropic domain thickness. ${\xi_0}\equiv 1$},
\label{fig1}
\end{figure}
\begin{figure}
\epsfxsize=10cm
\epsfbox{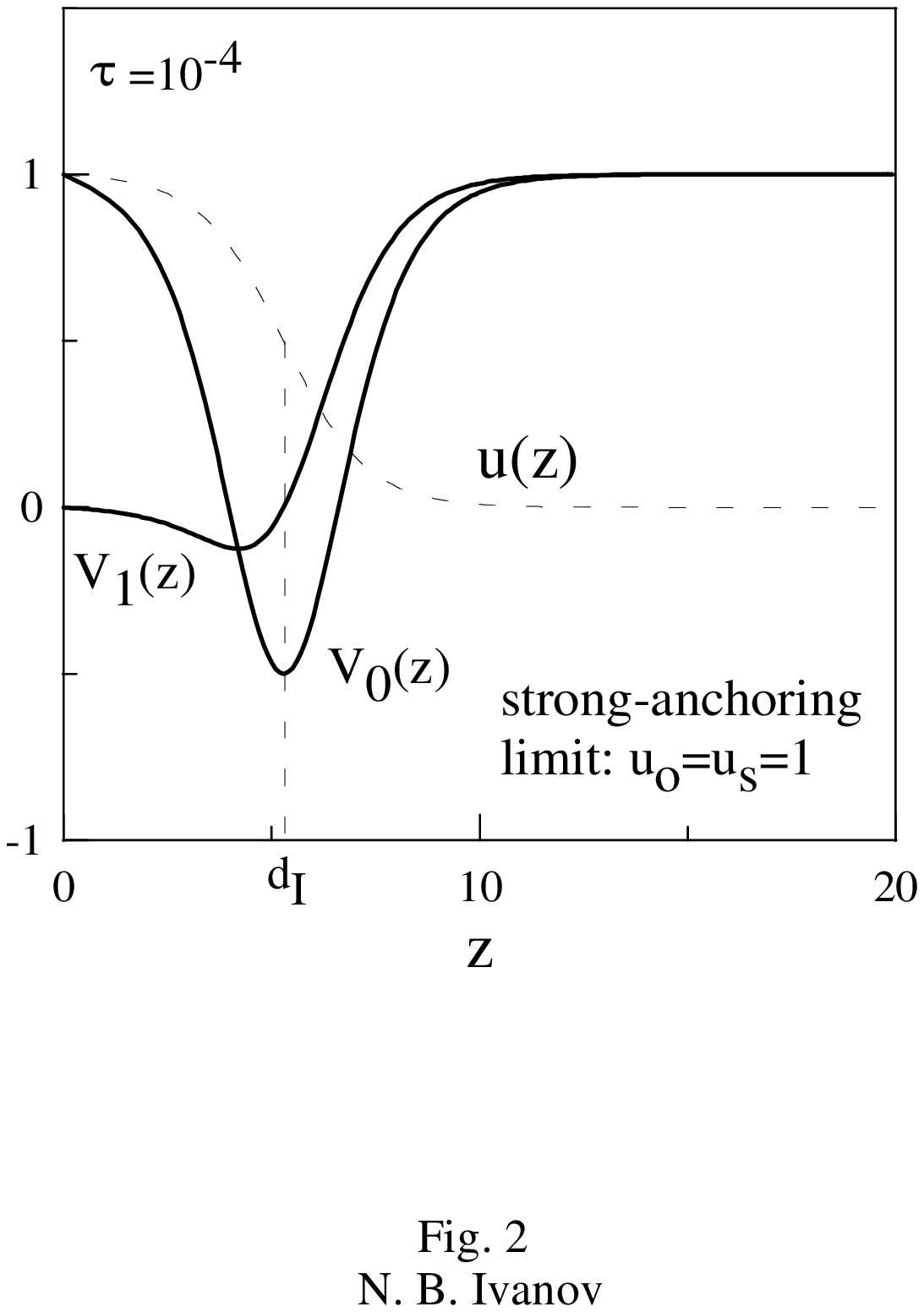}
\caption{The fluctuation mode potentials in Eq. (\ref{ne}) in
the srtong-anchoring limit. The wells in $V_0(z)$ and
$V_1(z)$ are connected with the scalar order-parameter
soft mode $\phi_0^{(0)}(z)$ and the bound uniaxial
director exciations $\phi_{1,2}^{(n)}(z)$, respectively.
${\xi_0}\equiv 1$}.
\label{fig2}
\end{figure}
\begin{figure}
\epsfxsize=10cm
\epsfbox{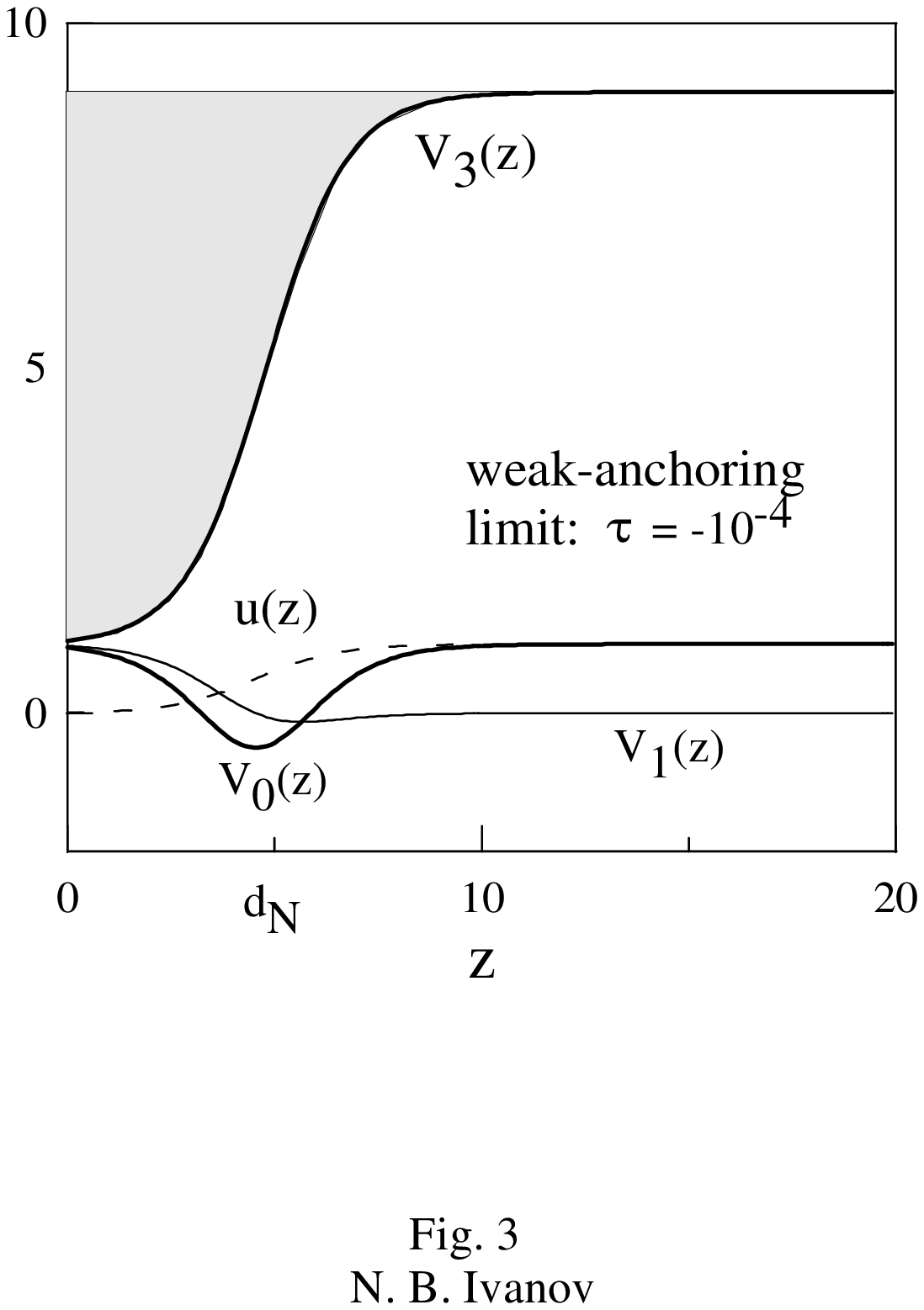}
\caption{The fluctuation mode potentials in Eq. (\ref{ne}) in
the weak-anchoring limit. The wells in $V_0(z)$ and
$V_3(z)$ (the shaded area) are connected with
the scalar order-parameter
soft mode $\phi_0^{(0)}(z)$ and the bound biaxial
director exciations $\phi_{3,4}^{(n)}(z)$, respectively.
${\xi_0}\equiv 1$}.
\label{fig3}
\end{figure}

\begin{references}
\bibitem{note3}
Here it is supposed that the variation of $\theta(z)$
throughout the interface is small: see, e.g.,  Ref.
\onlinecite{barbero}.
\end{references}
\end{document}